\documentstyle[aps,twocolumn,epsf]{revtex}

\newcommand{\nc}{\newcommand}
\nc{\renc}{\renewcommand}
\renc{\baselinestretch}{1.1}
\nc{\com}[1]{\ \\{\bf {\# #1}}}

\def\bort#1{ }
\newlength{\overeqskip}
\newlength{\undereqskip}
\nc{\be}[1]{\begin{equation} \mbox{$\label{#1}$}}
\nc{\bea}[1]{\begin{eqnarray} \mbox{$\label{#1}$}}
\nc{\Section}[2]{\section{\sc #2}\label{#1}\seqnoll}
\nc{\Subsection}[2]{\subsection{\sc #2}\label{#1}}
\nc{\Bibitem}[1]{\bibitem{#1}}
\nc{\Label}[1]{\label{#1}}
\nc{\eea}{\vspace{\undereqskip}\end{eqnarray}}
\nc{\ee}{\vspace{\undereqskip}\end{equation}}
\nc{\bdm}{\begin{displaymath}}
\nc{\edm}{\end{displaymath}}
\nc{\dpsty}{\displaystyle}
\nc{\bc}{\begin{center}}
\nc{\ec}{\end{center}}
\nc{\ba}{\begin{array}}
\nc{\ea}{\end{array}}
\nc{\bab}{\begin{abstract}}
\nc{\eab}{\end{abstract}}
\nc{\btab}{\begin{tabular}}
\nc{\etab}{\end{tabular}}
\nc{\bit}{\begin{itemize}}
\nc{\eit}{\end{itemize}}
\nc{\ben}{\begin{enumerate}}
\nc{\een}{\end{enumerate}}
\nc{\bfig}{\begin{figure}}
\nc{\efig}{\end{figure}}
\nc{\seqnoll}{\setcounter{equation}{0}}
\renc{\theequation}{\thesection.\arabic{equation}}
\nc{\refc}[1]{\mbox{Ref.~\cite{#1}}}
\nc{\refs}[1]{\mbox{Refs.~\cite{#1}}}
\nc{\eqs}[2]{\mbox{Eqs.~(\ref{#1}) and (\ref{#2})}}
\nc{\eq}[1]{\mbox{Eq.~(\ref{#1})}}
\nc{\figs}[2]{\mbox{Figs.~\ref{#1} and \ref{#2}}}
\nc{\fig}[1]{\mbox{Fig.~\ref{#1}}}

\nc{\figcap}[1]{\begin{quote}\refstepcounter{figure}
        {\bf Figure \thefigure}: {\small #1}\end{quote}}
\nc{\tabcap}[1]{\begin{quote}\refstepcounter{table}
        {\bf Table \thetable}: {\small #1}\end{quote}}

\nc{\tag}[1]{\label{#1} \marginpar{{\footnotesize #1}}}
\nc{\mtag}[1]{\label{#1} \mbox{\marginpar{{\footnotesize #1}}}}
\nc{\etal}{\mbox{\it et al. }}
\nc{\ie}{{\rm i.e. }}
\nc{\eg}{{\it e.g. }}
\nc{\arreq}{&\!\!\!=\!\!\!&}
\nc{\arrmi}{&\!\!\!!-\!\!\!&}
\nc{\arrpl}{&\!\!\!+\!\!\!&}
\nc{\arrap}{&\!\!\!\approx\!\!\!&}
\nc{\non}{\nonumber}
\nc{\nn}{\nonumber\\}
\nc{\align}{&&}

\nc{\mat}[4]{{\left(\ba{cc} #1 & #2 \\ #3 & #4 \ea\right)}}

\def\simleq{\; \raise0.3ex\hbox{$<$\kern-0.75em
      \raise-1.1ex\hbox{$\sim$}}\; }
\def\simgeq{\; \raise0.3ex\hbox{$>$\kern-0.75em
      \raise-1.1ex\hbox{$\sim$}}\; }

\nc{\DOT}{\hspace{-0.08in}{\bf .}\hspace{0.1in}}
\nc{\Laada}{\hbox {$\sqcap$ \kern -1em $\sqcup$}}
\nc\loota{{\scriptstyle\sqcap\kern-0.55em\hbox{$\scriptstyle\sqcup$}}}
\nc\Loota{{\sqcap\kern-0.65em\hbox{$\sqcup$}}}
\nc\laada{\Loota}
\nc{\qed}{\hskip 3em \hbox{\BOX} \vskip 2ex}
\def\Re{{\rm Re}\hskip2pt}
\def\Im{{\rm Im}\hskip2pt}
\nc{\real}{{\rm I \! R}}
\nc{\Z}{{\sf Z \!\!\! Z}}
\nc{\complex}{{\rm C\!\!\! {\sf I}\,\,}}
\def\diag{{\rm diag}}
\def\bigid{\leavevmode\hbox{\small1\kern-3.8pt\normalsize1}}
\def\id{\leavevmode\hbox{\small1\kern-3.3pt\normalsize1}}
\nc{\slask}{\hspace{0.1em}\not\hspace{-0.25em}}
\nc{\bis}{{\prime\prime}}
\nc{\pa}{\partial}
\nc{\na}{\nabla}
\def\>{\rangle}
\def\<{\langle}
\def\para{\parallel}
\nc{\goto}{\rightarrow}
\nc{\swap}{\leftrightarrow}
\nc{\EE}[1]{ \mbox{$\times 10^{#1}$} }
\nc{\abs}[1]{\left|#1\right|}
\nc{\at}[2]{\left.#1\right|_{#2}}
\nc{\norm}[1]{\|#1\|}
\nc{\abscut}[2]{\abs{#1}_{\scriptscriptstyle#2}}
\nc{\vek}[1]{\hbox{\boldmath$#1$}}
\nc{\integral}[2]{\int\limits_{#1}^{#2}}
\def\inv#1{\frac{1}{#1}}
\nc{\dd}[2]{{{\partial #1}\over{\partial #2}}}
\nc{\ddd}[2]{{{{\partial}^2 #1}\over{\partial {#2}^2}}}
\nc{\dddd}[3]{{{{\partial}^2 #1}\over
        {\partial #2 \partial #3}}}
\nc{\dder}[2]{{{d #1}\over{d #2}}}
\nc{\ddder}[2]{{{d^2 #1}\over{d {#2}^2}}}
\nc{\dddder}[3]{{d^2 #1}\over
        {d #2 d #3}}
\def\d#1#2{d\,^{#1}#2\,}
\def\dbar#1#2{\frac{d\,^{#1}#2}{(2\pi)^{#1}}}

\nc{\cc}{\mbox{$c.c.$ }}
\nc{\hc}{\mbox{$h.c.$ }}
\nc{\cf}{cf.\ }
\nc{\erfc}{{\rm erfc}}
\nc{\Tr}{{\rm Tr\,}}
\nc{\tr}{{\rm tr\,}}
\nc{\pol}{{\rm pol}}
\nc{\sign}{{\rm sign}}
\nc{\bfT}{{\bf T }}

\nc{\cA}{{\cal A}}
\nc{\cB}{{\cal B}}
\nc{\cD}{{\cal D}}
\nc{\cE}{{\cal E}}
\nc{\cF}{{\cal F}}
\nc{\cG}{{\cal G}}
\nc{\cH}{{\cal H}}
\nc{\cL}{{\cal L}}
\nc{\cM}{{\cal M}}
\nc{\cO}{{\cal O}}
\nc{\cP}{{\cal P}}
\nc{\cQ}{{\cal Q}}
\nc{\cT}{{\cal T}}
\nc{\1}{\aa}
\nc{\2}{\"{a}}
\nc{\3}{\"{o}}
\nc{\4}{\AA}
\nc{\5}{\"{A}}
\nc{\6}{\"{O}}
\nc{\al}{\alpha}
\nc{\Del}{\Delta}
\nc{\e}{\epsilon}
\nc{\eps}{\epsilon}
\nc{\g}{\gamma}
\nc{\lam}{\lambda}
\nc{\om}{\omega}
\nc{\Om}{\Omega}
\nc{\ve}{\varepsilon}
\nc{\mn}{{\mu\nu}}
\nc{\ka}{\kappa}

\nc{\vp}{\varphi}
\nc{\pub}[4]{\Bibitem{#1}#2, {\sl ``#3''}, #4.}

\def\ijmp#1#2#3{{\it Int.\ J.\ Mod.\ Phys.\ }{{\bf #1} {(#2)} {#3}}}

\def\np#1#2#3{{\it  Nucl.\ Phys.\ }{{\bf #1} {(#2)} {#3}}}
\def\npfs#1#2#3#4{{\it  Nucl.\ Phys.\ }{{\bf #1 [FS#2]} {(#3)} {#4}}}

\def\pl#1#2#3{{\it  Phys.\ Lett.\ }{{\bf #1} {(#2)} {#3}}}

\def\rpp#1#2#3{{\it Rep.\ Prog.\ Phys.\ }{{\bf #1} {(#2)} {#3}}}

\def\vk{{\bf k}}

\begin{document}

\title{Stability of large scale chromomagnetic fields
	in the early universe}

\author{Per Elmfors\raisebox{1ex}{a}
 and David Persson\raisebox{1ex}{b}}

\address{\raisebox{1ex}{a}Department of Physics,
   Stockholm University,
 S-113 85 Stockholm, Sweden\\
elmfors@physto.se\\
\raisebox{1ex}{b}Department of Physics, University of British Columbia,
        6224 Agricultural Road, Vancouver, B.C. V6T 1Z1, Canada
persson@physics.ubc.ca}

\maketitle

\begin{abstract}
It is well known that Yang-Mills theory in vacuum has a perturbative
instability to spontaneously form a large scale magnetic field (the Savvidy
mechanism) and that a constant field is unstable so that a possible ground
state has to be inhomogenous over the non-perturbative scale $\Lambda$ (the
Copenhagen vacuum).
We argue that this spontaneous instability does not occur at high temperature
when the induced field strength
$gB\sim \Lambda^2$ is much weaker than the magnetic mass squared $(g^2T)^2$.
At high temperature oscillations of gauge fields acquire a thermal
mass  $\cM\sim gT$
and we show that this mass stabilizes a magnetic field
which is constant over length scales shorter than the magnetic screening
length $(g^2T)^{-1}$. We therefore
conclude that there is no indication for any
spontaneous generation
of weak non-abelian magnetic fields in the early universe.
\end{abstract}


\narrowtext
\begin{center}
  \Section{s:intro}{Introduction}
\end{center}
The perturbative ground state in non-abelian gauge theory is unstable
towards a generation of a background field. This was first discovered by
Savvidy \cite{Savvidy77} who calculated the effective potential in a
uniform background magnetic
field and found a non-trivial minimum. Shortly after Nielsen and
Olesen \cite{NielsenO78} noticed that this minimum is unstable and this
discovery 
stimulated a whole series of papers about how a non-uniform background field
develops and stabilizes the system \cite{AmbjornNO79}.

One question that has been asked many times is whether this instability
occurs also at high temperature. The  gauge bosons in
the electroweak model have large masses generated by the Higgs mechanism so
they screen any $SU(2)$ magnetic field on very short scales below the critical
temperature. At high temperature  QCD can
to some extent be described by perturbation theory around
the trivial vacuum. To find
observable effects from a magnetic instability it is therefore motivated to
go to the very early universe. In particular it was argued in
\cite{EnqvistO94} that a ferromagnetic phase of Yang-Mills theory could
possibly be responsible for generating the seed field that is
needed in many models to explain the observed intergalactic magnetic field
\cite{Kronberg94}.

In this article we are going to argue that no such instability occurred in
the early universe. The simple argument is that the typical long wavelength
modes that are responsible for the instability in the first place have
typically a much longer correlation length than the magnetic screening
length. Therefore, these modes do not see a uniform background field but
only statistical fluctuations and the whole mechanism for the instability is
not present. We furthermore derive the dispersion relation for
excitations in the presence of a constant magnetic field (typically
much stronger than
the one generated by the Savvidy mechanism) and we find that there are stable
modes for fields much weaker than the plasma mass, while for strong
fields the situation is much less clear. These background fields are however
not generated {\em spontaneously}.
The reason for the stability is simply that the
plasma generates a mass for all propagating modes at long wavelengths.

\begin{center}
  \Section{s:screening}{The instability and how it is screened}
\end{center}
A constant field strength is not a gauge invariant concept for non-abelian
field theories but one can define classes of gauge equivalent gauge
potentials
that can be transformed into a given
constant field strength by a gauge transformation.
It has been shown \cite{BrownW79}
that there are actually two gauge-inequivalent classes. Let
us for simplicity consider $SU(2)$ Yang-Mills theory throughout the paper
and give examples of the two classes. For a constant
magnetic field $F^3_{xy}=-B$, a representative of the
first class is
\be{nonabgauge}
	A^1_x=\sqrt{\frac{B}{g}}~,\quad A^2_y=-\sqrt{\frac{B}{g}}~~,
\ee
where the field strength is generated by the non-trivial commutation
relations of the different components in \eq{nonabgauge}.
This field strength is
not covariantly conserved and requires therefore a uniform source to fulfill
the equations of motion. Since we do not know of any reason to expect such a
coherent source over large scales in the early universe we shall instead
concentrate on the other class of gauge fields which are the close analogue
of the usual abelian magnetic field
\be{abgauge}
	A^a_\mu=\delta^{a3}(0,0,-Bx,0)~~,
\ee
where the field points in the 3-direction in group space.
Although this field is gauge dependent we only calculate
quantities that are gauge independent so we can safely choose a convenient
gauge as above.

The energy spectrum of charged vector
fluctuations in the background field in \eq{abgauge} is
\be{spect}
	E^2(k_z,l,\sigma)=k_z^2+(2l+1)gB-2\sigma gB~~,
\ee
where the term  $(2l+1)gB$, $l=0,1,2,\ldots$,
comes from the orbital angular momentum and the term $2\sigma
gB$, $\sigma=\pm 1$, is the spin energy. For spin one particles the spin
coupling to the magnetic field overcompensates the cost of confining the
excitation in the plane orthogonal to the magnetic field and this is why there
is an unstable mode for $k_z^2<gB$. The real part of the
effective potential at zero
temperature for a $SU(N)$ theory \cite{Savvidy77}
\be{vacEP}
	\Re V(B)=\inv 2B^2+\frac{11N}{96\pi^2}g^2B^2
	\left(\ln\frac{gB}{\mu^2}-\inv2\right)~~,
\ee
has a minimum at the renormalization group invariant scale
\be{gBmin}
	gB_{\rm min}=\Lambda^2
	=\mu^2\exp\left(-\frac{48\pi^2}{11Ng^2(\mu)}\right)~~.
\ee
There is also an imaginary part of the effective potential with the
interpretation that a constant background field is unstable and decays to
a non-uniform configuration \cite{AmbjornNO79}.
The length scale over which the magnetic field varies is given by the only
dimensionful constant in the problem, namely $\Lambda$.
\Subsection{ss:scales}{Scales}
Like in all other problems in physics we need first to sort out the
different scales that enter into the problem and see if there is any
hierarchy between them.
\bit
\item[$T$:]
The highest energy scale is given by the temperature for the situations we
have in mind. This is the typical energy of particles in the plasma and the
inter-particle distance is $\sim 1/T$.
\item[$gT$:]
The interaction of soft particles ($p\sim gT$) with hard particles ($p\sim
T$) generates a thermal mass of order $gT$. Though static magnetic
correlation functions are unscreened by this mechanism all propagating modes
pick up a plasma mass of this order.
\item[$g^2T$:]
On this momentum scale Yang-Mills theory becomes non-perturbative. All
diagrams contribute to the same order. We can therefore not analytically
calculate correlation functions on too large length scale. On the other
hand, lattice simulations and arguments 
from dimensional reduction show that non-abelian magnetic
fields are screened over the length scale $1/g^2T$. Thus, no magnetic fields
can be constant on this scale.
\item[$\Lambda$:]
The strong coupling scale below which the vacuum theory becomes
non-perturbative depends on the particular theory we have in mind. For QCD we
have $\Lambda_{\rm QCD}\simeq 200$~MeV. For an $SU(2)$ theory with a
coupling constant $g(\mu)=0.65$ at $\mu=M_Z=91$~GeV the scale is
$\Lambda_{SU(2)} =8\EE{-10}$~GeV.
A hypothetical $SU(5)$ GUT with $g(\mu)=0.7$ at
$\mu=10^{15}$~GeV would have $\Lambda_{SU(5)}=2\EE{-4}\mu=2\EE{11}$~GeV.
\eit
In order to possibly have an instability the symmetry has to be unbroken and
so the temperature has to be much higher than $\Lambda$ for the examples we
give above. In this paper we only consider $SU(N)$ explicitly (and mostly
only $SU(2)$) and in that case one can derive the bound $g(T)^2TN>58
\Lambda$ from \eq{gBmin}. Therefore, we
shall assume that we have a clear scale separation with the hierarchy
\be{hier}
	\Lambda\ll g^2T\ll gT\ll T~~.
\ee
Now we have to put the in the scale $gB$ and we are going to discuss two
possibilities.
\Subsection{ss:savscr}{Screening of the Savvidy instability}
In the Savvidy mechanism for
generating a magnetic field the magnitude is given in \eq{gBmin} and is thus
much weaker than any thermal scale. One should then remember that in  order
to derive \eq{vacEP} a constant field was assumed and it was the modes with
momentum $k_z^2<gB,~\<k_x^2+k_y^2\>=gB$
that gave rise to both the instability and the imaginary part
at the non-trivial minimum. These modes have a spatial extension that is of
the order $1/\sqrt{gB}$ and thus much larger than the magnetic screening
length $1/g^2T$. Even though it has not been possible to rigorously
calculate the magnetic screening length analytically there is overwhelming
numerical evidence \cite{BilloireLS81},
as well as theoretical arguments \cite{Linde80},
that non-abelian magnetic fields are screened over the distance
$1/g^2T$. The modes that are responsible for the instability do therefore not
experience a constant field but only random thermal
fluctuations. Consequently, there are no indications that any large scale
magnetic field is present in a high temperature Yang-Mills gas. It should
however be kept in mind that the physics of such a gas on large length scales
is very poorly known and it is difficult to make quantitative statements.
\Subsection{ss:extmag}{External magnetic fields}
Even if there is no spontaneous generation of magnetic fields it is still
interesting to study the properties of a  field on scales shorter
than the screening length. Such a field could be generated in first order
phase transitions and in cosmic strings. The smoothness of such fields
depends on the exact situation but we shall concentrate on fields
which are constant over
the size of the lowest Landau level, \ie $L\sim 1/\sqrt{gB}$.  For very
strong fields  we expect that thermal effects do not matter and
we would end up in the unstable situation just like in the vacuum.
To start from such a state and try to add thermal corrections is bound to be
difficult since the starting point is unstable \cite{MaianiMRT86}. We shall
therefore start from the stable high temperature phase and turn on a weak
external field.
The relevant scale to compare with is the thermal mass $gT$ of the propagating
modes. Therefore, we shall assume that the field is stronger than the
magnetic scale in order to be constant over the size of the lowest Landau
level but still weaker than $(gT)^2$. The hierarchy we end up with is thus
\be{bhier}
	\Lambda^2\ll (g^2T)^2\ll gB\ll (gT)^2\ll T^2~~.
\ee
This separation of scales assumes a small coupling constant which is only
marginally valid in many theories of interest but this is the only way of
making sense of perturbation theory. We now expect that for field in the
range of \eq{bhier} there are only small corrections to the stable
dispersion relations of propagating modes while as $gB\goto (gT)^2$ an
instability may develop.
\begin{center}
  \Section{s:se}{Self energy in a magnetic field}
\end{center}
We shall now calculate the self-energy in a background magnetic field at
high temperature and see how the dispersion relations are affected.
As a concrete example we study a $SU(2)$ gauge theory with a background
field given by \eq{abgauge}.
The Lagrangian in a background field is given by
\bea{Lag}
	\cL&=&-\inv4 F^2-\inv{2}\left((\pa_\mu Z_\nu)^2-
	(1-\inv{\xi})(\pa_\mu Z^\mu)^2\right)
	\nn &&
	-\left(|D_\mu W_\nu|^2-(1-\inv{\xi})|D_\mu W^\mu|^2\right)
	\nn &&
	-igF^{\mu\nu}W_\mu^\dagger W_\nu
	-ig\pa_\mu Z_\nu (W_\mu^\dagger W_\nu - W_\nu^\dagger W_\mu)
	\nn &&
	-ig (Z_\mu W_\nu^\dagger-Z_\nu W_\mu^\dagger) D_\mu W_\nu
	\nn &&
	-ig (D_\nu W_\mu)^\dagger (Z_\mu W_\nu-Z_\nu W_\mu)
	\nn &&
	-\frac{g^2}{2}\left((W^\dagger\cdot W)^2
	-|W\cdot W|^2
	\right. \nn &&\quad \left.
	+2 Z^2 W^\dagger\cdot W-2 |Z\cdot W|^2 \right)
	\nn &&
	+ ~{\rm ghost~terms.}
\eea
The background field points in the 3-direction in group space and we use the
notation $Z_\mu=W_\mu^3$ and $W_\mu=(W^1_\mu+iW^2_\mu)/\sqrt{2}$.
The diagrams that contribute to the polarization tensor at one loop are the
same as usual but with propagators in a background field and the 3-point
vertex with derivatives has covariant derivatives with respect to the
charged particle lines. We shall use the Feynman gauge $\xi=1$.
The $W$-boson propagator in the Schwinger
representation is then
\bea{Wprop}
	G^W_{\mu\nu}(x,x')&=&\phi(x,x')\int \dbar4P e^{-iP(x-x')}
	G^W_{\mu\nu}(P)
	\nn
	G^W_{\mu\nu}(P)&=&-\int_0^\infty \frac{ds}{\cos (gBs)}
	(\exp[-2s gF])_{\mu\nu}
	\nn &&\times \exp[is(P_\para^2
	+\frac{\tan gB s}{gB s}P_\perp^2)]~~,
\eea
where the gauge dependence is contained in the phase factor
\be{phiexpl}
	\phi(x,x')=\exp\{ig \int_{x'}^x d \xi^\mu [A_\mu+\inv2 F_{\mu\nu}
	(\xi-x')^\nu ]\} ~~,
\ee
the rest being both  translational invariant and invariant under
abelian gauge transformations generated by $U=e^{i\alpha(x) \sigma^3}$.
The  exponential of the field strength can be expanded as
\bea{eF}
	(\exp[-2s gF])_{\mu\nu}&=&g_{\mu\nu}-gF_{\mu\nu}\frac{\sin(2gBs)}{gB}
	\nn &&
	-\frac{(F^2)_{\mu\nu}}{(gB)^2}(1-\cos(2gBs))~~.
\eea
In  the bubble diagram it is convenient to integrate the 3-point
vertex derivatives
by parts so they only act on the propagators inside the loop. Then one can
use the rule
\bea{DG}
	&&(i\pa_\mu+gA_\mu)G^W_{\alpha\beta}(x,x')
	\nn
	&=&\phi(x,x')\int \dbar4P e^{-i P(x-x')}
	\nn &&\times
	\left(-\frac{i}{2}gF_{\mu\nu}\frac{\pa}{\pa P_\nu}
	+P_\mu\right) G^W_{\alpha\beta}(P)~~>
\eea
It follows from the form of the propagators that the polarization tensor can
be written as
\be{phiPi}
	\Pi_{\mu\nu}(x,x')=\phi(x,x')\int \dbar4P e^{-iP(x-x')}
	\Pi_{\mu\nu}(P)~~,
\ee
where again the gauge dependence is only in phase factor.

One problem we encounter immediately if we try to do a strict perturbative
expansion in the background field is that the Landau level
solutions in a background field cannot be approximated by small
perturbations from a plane wave basis. We shall therefore use Landau levels
$W_\mu(\kappa,x)$
as external states and calculate the polarization tensor for small
field strengths. Another concern is gauge invariance of the external fields
and the phase factor in \eq{phiPi}. We shall see in Section~\ref{s:screenLL}
that when calculating expectation values of the form
\bea{DRvev}
	&&\int d^4x\,d^4x'\,W_\mu^\dagger(\kappa,x)
	\nn &&\times
	\left[-\delta(x-x')
	(g^{\mu\nu}D^2-D^\mu D^\nu
	-2igF^{\mu\nu})
	\right. \nn && \left.
	\quad-\Pi^{\mu\nu}(x,x')\right]W_\nu(\kappa',x')
\eea
the phase factor in $\Pi_{\mu\nu}$ combine with the external Landau levels
to give a factor $\delta(\kappa-\kappa')$, where $\kappa=(k_0,k_z,k_y,n)$ are
the quantum numbers of the Landau levels.
We shall therefore now
calculate $\Pi_{\mu\nu}(P)$ for weak fields and evaluate the expectation values
in Section~\ref{s:screenLL}.

As pointed out above, the diagrams to one-loop are the standard ones except
that the full $W$-propagators and covariant derivatives should be used. It is
then trivial to see that at lowest order the standard hard thermal loop
result is recovered (assuming $T\gg K$)
\bea{HTLPi}
	\Pi(K)_{\mu\nu}=\frac{3\cM^2}{2}\int\frac{d\Omega}{4\pi}
	\biggl[\align g_{\mu\nu}-\frac{K_\mu u_\nu+u_\mu K_\nu}{u\cdot K}
	\nn \align
	+\frac{u_\mu u_\nu}{(u\cdot K)^2}\biggr]~~,
\eea
where $\cM^2=\frac{g^2NT^2}{9}$, $u_\mu=(1,\vec{u})$
and the integral is over the angles of $\vec{u}$.
The first order correction in $gB$ is, not surprisingly, IR
divergent.
The full calculation is difficult to perform, but for small loop
momenta, and  assuming that there is
regularizing mass $m$, one
can show that correction to $\Pi(K)$ goes like
\be{pismall}
	g^2 T \left[ \sqrt{m^2+2gB}-\sqrt{m^2-2gB} \right]~~~.
\ee
If the thermal gluon mass on the electric scale $m \sim gT$ is the
regularizing scale, as suggested in \cite{CaboKS81}, this gives
\be{pielmass}
	\Delta\Pi \sim g^2 T gB/m \sim g\, gB \ll gB~~,
\ee
which is much smaller than the tree level $gB$.
However, static magnetic modes are not screened in the HTL
resummation scheme.
If the mass is instead  generated on the magnetic scale $m \sim g^2 T$,
with our assumption of a clear scale separation with
$g^2T\ll \sqrt{gB}$, the instability is instead regularized by the field
itself
\be{pimagmass}
	\Delta\Pi \sim g^2 T \sqrt{gB} (1+i) \ll gB~~~,
\ee
and again smaller than the  tree level $gB$. An imaginary part appears here due
to the unstable mode on tree level, but it is of the same order as the
real part, and thus  negligible.
Furthermore, we find that the contributions linear in $B$ from hard internal
momenta are suppressed compared to the tree level $gB$.
We can therefore
conclude that it is enough to consider the leading HTL term for
$\Pi(K)$.

It may now seem that we have approximated away all non-trivial effects from
having a  heat-bath {\em and} a magnetic field.  This is however not so
because when we calculate the expectation value of $\Pi_{\mu\nu}(K)$ to find
the mass gap the orthogonal momentum is $\<k_x^2+k_y^2\>=(2l+1)gB$, where 
$l=0,1,2, \ldots$
is the orbital angular momentum quantum number.
Therefore the
thermal mass is shifted by an amount $g^2T^2\<k^2/k_0^2\>\sim
g^2T^2\,gB/(g^2T^2)\sim gB$. The fact that
even the lowest Landau level is localized makes a shift through the
self-energy that is of the same order as the tree level term and is
therefore the dominant correction.

\begin{center}
  \Section{s:screenLL}{Screening  in the lowest Landau level}
\end{center}
The equation of motion for charged YM fields in a background of a constant
chromo-magnetic field is given by
\be{eqom}
	D^2 W_\mu-D_\mu D\cdot W-2igF_\mu^{~\nu}W_\nu=0~~,
\ee
where $F_\mu^{~\nu}$ is the background field strength and
$D_\mu=\pa_\mu-igA_\mu$ is the background
covariant derivative. $W_0$ is not a
physical degree of freedom but it is algebraically related to the currents 
and the gauge is fixed using
the condition $D\cdot W=0$ as usual. For the general solution of
\eq{eqom} in the gauge $A_\mu=(0,0,-Bx,0)$ we use the Ansatz
\bea{ansatz}
	&& W_\mu(k_0,k_z,k_y,n;x)\equiv\sum_a c_a W_\mu^{(a)}(\kappa;x)
	\nn &&
	=\sum_a c_a\, w^{(a)}_\mu
	\exp[-i(k_0x_0-k_z z-k_y y)] I_{n_a,k_y}(x)~,
\eea
where $I_{n_a,k_y}(x)$ can be expressed in terms of Hermite polynomials as
\bea{I}
	I_{n,k_y}(x)&=&\left(\frac{gB}{\pi}\right)^{1/4}
	\exp\left[-\frac{gB}{2}(x+\frac{k_y}{gB})^2\right]
	\nn &&\times
	\inv{\sqrt{n!}} H_n\left[\sqrt{2gB}(x+\frac{k_y}{gB})\right]~~.
\eea
For the vector structure of $W_\mu$ we choose the basis
\bea{wbasis}
	&&w^{(0)\mu}=(1,0,0,0)\nn
	&&w^{(z)\mu}=(0,0,0,1)\nn
	&&w^{(+)\mu}=\frac{1}{\sqrt{2}}(0,1,i,0)\\
	&&w^{(-)\mu}=\frac{1}{\sqrt{2}}(0,1,-i,0)\non~~,
\eea
and they are normalized to $w^{(a)\dagger}_\mu w^{(b)\mu}=g^{ab}=
\diag(1,-1,-1,-1)$.
They  have the convenient properties
\be{Fw}
	F_\mu^{~~\nu} w^{(0,z)}_\nu=0~,\quad
	F_\mu^{~~\nu} w^{(\pm)}_\nu=\pm iB w^{(\pm)}_\mu~~.
\ee
Energy eigenfunctions are then given by \eq{ansatz}
with $a\in\{0,z,+,-\}$, $n_0=n_z=n-1$, $n_+=n-2$ and $n_-=n$. We use
the convention that $I_{-1}=I_{-2}=0$, or equivalently that $c_z=c_+=0$
for $n=0$ and $c_+=0$ for $n=1$. For this Ansatz \eq{eqom}
becomes
\bea{eveq}
	&&(D^2 g_\mu^{~\nu}-2ig F_\mu^{~\nu})W_\nu(\kappa;x)
	\nn &&
	=(-k_0^2+k_z^2+(2n-1)gB)W_\mu(\kappa;x)~~.
\eea
The gauge condition $D\cdot W=0$ should now be imposed on this solution and
that gives us the constraint
\be{cconstraint}
	c_0k_0-c_zk_z -i c_+\sqrt{gB(n-1)} +i c_-\sqrt{gBn}=0~~.
\ee
For the evaluation of \eq{DRvev} the essential integrals are of the form
\bea{xint}
	&&\int \d4x\,\d4x'\,W^{(a)\dagger}_\mu(\kappa,x)
	\Pi^{\mu\nu}(x,x')W^{(b)}_\nu(\kappa',x')
	\nn&=&
	\int \dbar4P \d4x\d4x'
	w^{(i)\dagger}_\mu \Pi^{\mu\nu}(P) w^{(j)}_\nu
	e^{-iP_\alpha(x-x')^\alpha}\phi(x,x')
	\nn &&\times
	\exp[i(k_0x_0-k_zz-k_yy)-i(k_0'x_0'-k_z'z'-k_y'y')]
	\nn &&\times
	I_{n_a}(x) I_{n_b'}(x')
	\nn&=&
	(2\pi)^3\delta^{(0,z,y)}(\kappa-\kappa')\int_0^\infty dp_\perp p_\perp
	w^{(a)\dagger}_\mu \Pi^{\mu\nu}(k_0,k_z,p_\perp)
	\nn &&\times
	 w^{(b)}_\nu
	\exp\left[-\frac{p_\perp^2}{gB}\right]\nn
	&&\times
	2^{(n_b'-n_a)/2}(-1)^{n_b'}\sqrt{\frac{n_a!}{n_b'!}}
	\left(\frac{i(p_x+ip_y)}{\sqrt{gB}}\right)^{n_b'-n_a}
	\nn &&\times
	\frac{2}{gB}L^{n_b'-n_a}_{n_a}\left(\frac{2p_\perp^2}{gB}\right)~~.
\eea
The factor $w^{(a)\dagger}_\mu \Pi^{\mu\nu}(k_0,k_y,p_\perp) w^{(b)}_\nu$
depends only on rotational invariant quantities apart from factors of
$p_x\pm ip_y$. The angular integral over $\vec{p}_\perp$ then constrains
the total power of $p_x\pm ip_y$ to zero. This gives a $\delta$-function
between $n_a$ and $n_b'$ such that, depending on $a$ and $b$, different
energy levels have vanishing overlap.

In particular we can study the polarization tensor in the lowest Landau level
where we only have the polarization state $w^{(-)}_\mu$ which automatically
fulfills the gauge condition. The dispersion relation takes the form
\bea{LLLPi}
	k_0^2+gB-k_z^2+\align\int_0^\infty dp_\perp\,\frac{2p_\perp}{gB}
	e^{-\frac{p_\perp^2}{gB}}
	\nn &&\times
	w^{(-)\dagger}_\mu \Pi^{\mu\nu}(k_0,k_z,p_\perp) w^{(-)}_\nu
	=0~~.
\eea
It is straightforward to expand $w^{(-)\dagger}_\mu
\Pi^{\mu\nu}(k_0,k_z,p_\perp) w^{(-)}_\nu$ using
$\Pi_{\mu\nu}=\cP_{\mu\nu}\Pi_T+\cQ_{\mu\nu}\Pi_L$ where
\bea{PLPT}
	\cP_{ij}(K)&=&-\delta_{ij}+\frac{k_ik_j}{k^2}~~,
	\quad\cP_{0\mu}=0~~,\nn
	\cQ_{\mu\nu}(K)&=&-\inv{K^2k^2}
	(k^2,-k_0\vk)_\mu (k^2,-k_0\vk)_\nu~~.
\eea
{}From \eq{HTLPi} we have that
\be{PiTL}
\ba{rcl}\displaystyle
   \Pi_T(K)&=&\displaystyle\frac{3\cM^2}{2}
        \left[\frac{k_0^2}{k^2}+(1-\frac{k_0^2}{k^2})
        \frac{k_0}{2k} \ln\left(\frac{k_0+k}{k_0-k}\right)\right]
	~~, \\[5mm]
   \Pi_L(K)&=&\displaystyle 3\cM^2
        (1-\frac{k_0^2}{k^2})
        \left[1-\frac{k_0}{2k}
        \ln\left(\frac{k_0+k}{k_0-k}\right)\right]
	 ~~.
\ea
\ee
In the lowest Landau level, we thus get
\bea{wPiw}
	&& w^{(-)\dagger}_\mu
	\Pi^{\mu\nu}(k_0,k_z,p_\perp) w^{(-)}_\nu
	\nn &&
	=-\frac{2k_z^2+p_\perp^2}{2(k_z^2+p_\perp^2)} \Pi_T
	-\frac{k_0^2 p_\perp^2}{2(k_0^2-k_z^2-p_\perp^2)(k_z^2+p_\perp^2)}
	\Pi_L~~.\nn[-3mm]
\eea
It is interesting to notice that the
longitudinal polarization contributes
to this lowest mode. It is so because $w_\mu^{(-)}$ points in the plane
perpendicular to the magnetic field and the momentum cannot be zero in the
plane.
For the question of stability of this mode we are interested low
momenta so we put $k_z=0$
which in this context means that $k_z^2\ll gB$
while being larger than $g^2T$ since we cannot probe the magnetic scale.
The mass shell condition $k_0^2=\cM^2$ including
$\cO(gB)$ corrections then becomes
\be{LLLdr}
	k_0^2+gB-\cM^2\left(1+\frac{2gB}{5\cM^2}\right)=0~~,
\ee
that is $k_0\simeq \cM(1-3gB/10\cM^2)$. A weak field results in a small
reduction of the plasma mass but plasmons remains stable (up to higher order
in $g$). We notice that \eq{wPiw} vanishes for $k_0=0$ and it is natural to
wonder whether there is any new low energy mode that potentially becomes
unstable as the field is switched on. Expanding \eq{wPiw} in powers of $k_0$
we find the dispersion relation
\be{smallk0}
	k_0^2+gB+i\frac{3\pi^{3/2}}{8}\frac{k_0\cM^2}{\sqrt{gB}}=0~~.
\ee
There is no solution to this equation with $k_0^2\sim gB$ and the dominating
imaginary part comes from Landau damping. 
\ie scattering with particles in the heat
bath.

As the field is increased the tree level $gB$ becomes comparable to the
plasma mass $\cM^2$ and one could expect the instability to reappear. The
analysis in Section~\ref{s:se} showed that \eq{LLLPi} contains the dominant
corrections as long as $g$ is small and $gB\gg (g^2T)^2$. It can therefore
be used all the way up to $gB\simleq \cM^2$, but the analysis in terms of
a dispersion relation breaks down
when the imaginary part of the self-energy becomes large.
The integration in \eq{LLLPi} goes
below the light cone ($k_0^2<k_z^2+p_\perp^2$) where the polarization tensor
has an imaginary part due to Landau damping and synchrotron absorption
(emission is not possible in the lowest Landau level). It increases with the
field and eventually the width of the excitation is so large
that it is not possible to talk about a quasi-particle.

\begin{figure}
\unitlength=1mm
\begin{picture}(100,105)(0,0)
\includegraphics{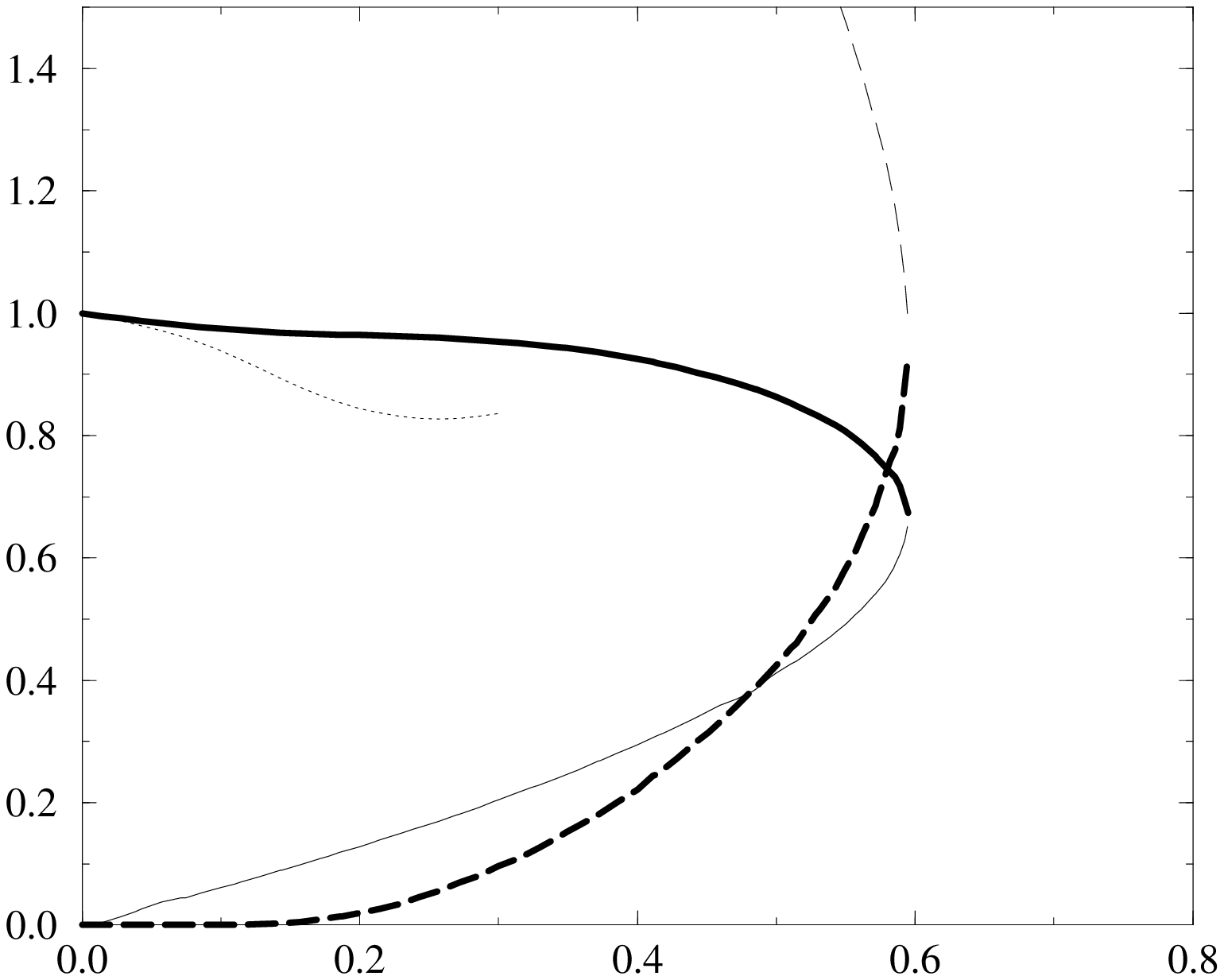}
   \put(68,1){$gB/\cM^2$}
   \put(30,65){$\dpsty\frac{k_0}{\cM}$}
   \put(45,73){$\dpsty\frac{-\Im \Pi}{k_0\cM}$}
   \put(25,48){$Z$}
\end{picture}
\figcap{The solid lines (fat and thin) show the solutions to \eq{LLLPi} for
$k_z=0$ using only the real part of the self-energy. There are two branches
of which the fat line corresponds to the standard plasmon. These two lines
are only meaningful if $-\Im\Pi/k_0$, is much smaller than
$k_0$. Therefore, we plot $-\Im\Pi/k_0$ for the two branches using dashed
lines. Only the fat line can satisfy the condition $-\Im\Pi\ll k_0^2$ for
small fields while the thin mode is simply an artefact of neglecting the
imaginary part of the self-energy. To further emphasize  that there is only
one propagating mode we plot the spectral weight of the fat mode (dotted
line), as calculated in \eq{Z}, and it shows that the upper mode saturates the
spectral weight for weak fields.
\label{f:dr}
}
\end{figure}

In order to estimate what happens as the field is increased
we have solved the dispersion relation in
\eq{LLLPi} numerically using only the real part. The result is shown in
\fig{f:dr}. In the same figure we show the imaginary part for the same value
of $gB$ and $k_0$. Our interpretation is that for weak fields ($gB\simleq
0.2\cM^2$) there is only one stable mode. For $gB\simgeq 0.2\cM^2$ the
imaginary part becomes so large that the spectral function looks more like a
broad resonance. Long before $gB$ reaches $\cM^2$ the imaginary part is so
large that  the quasi-particle language is meaningless. It should be
emphasized that the imaginary part we are discussing is {\em not}  the one
that indicates a spontaneous generation of a magnetic field, but simply the
Landau damping that attenuates all excitations. In particular, the sign of
the imaginary part corresponds to an exponentially damped mode.

To get further insight in what modes are present for weak fields we have
calculated the spectral weight of the  $k_0\simeq\cM$ mode using
\bea{Z}
	Z^{-1}(k_0,gB)&=&\inv{2k_0}\frac{d}{dk_0}
	\left(k_0^2+gB-\int d p_\perp
	\frac{2p_\perp }{gB}e^{-\frac{p_\perp^2}{gB}} 
	\right. \nn &&\times \left.
	w^{(-)\dagger}_\mu \Pi^{\mu\nu}(k_0,k_z,p_\perp) w^{(-)}_\nu\right) ~~,
\eea
and evaluate it on the solution to \eq{LLLPi}.
This formula makes sense only if the branch is a narrow resonance.
We have normalized $Z$ so that it
equals to  one for free a particle. In \fig{f:dr}  $Z$  is plotted for weak
fields and it is obvious that it is very close to one, leaving no spectral
weight to other modes.
To conclude: for weak fields there is only one propagating mode and it has a
large massgap $\simeq\cM$ while for increasing fields Landau damping
and synchrotron absorption rapidly damps all modes.
%

\begin{center}
  \Section{s:concl}{Conclusions}
\end{center}
As discussed in the introduction, rather soon after the non-trivial minimum
of the effective potential in vacuum was found, it was also realized that
this  minimum is unstable. Moreover, it was realized that the starting point
of a constant field is inconsistent and one is forced to consider
non-uniform background fields. The characteristic length scale on which the
background field has to vary is given by the renormalization scale
$\Lambda$. For QCD this would be $\Lambda_{\rm QCD}\simeq 200$~MeV while for
a $SU(2)$ theory $\Lambda_{SU(2)}\simeq 8\EE{-10}$~GeV if we take the values of
$g(\mu)$ from the $SU(2)$ sector of the electroweak theory. In a GUT
we made an estimate of $\Lambda_{SU(5)}\simeq 10^{11}$~GeV. As long as we
consider theories at temperature far above non-perturbative scale $\Lambda$
we would typically have that $\Lambda\ll g^2T$ and we do not expect the
perturbative instability to have any consequence for the high temperature
behaviour. In particular we do not expect that the free energy (or effective
potential) should have any non-trivial minimum. The non-trivial minimum in
vacuum comes entirely from the unstable mode. We have now found that this
mode is stabilized by the thermal heat bath and therefore we expect only a
trivial minimum. It is difficult  to say anything rigorous about
what happens with the background field on the scale $(g^2T)^{-1}$ and
larger, but since $\Lambda\ll g^2TN$ and magnetic fields are expected to be
screened by the mass $g^2 TN$ we do not find any support for the idea that a
nontrivial field configuration should be generated {\em spontaneously} in the
early universe.\medskip

The situation is very different if we consider an external
magnetic field that can be tuned at will. For weak fields the lowest
Landau level has a weakly damped massive plasma mode but it
acquires a large imaginary part from Landau damping for $gB\simgeq 0.2\cM^2$.
The equilibrium configuration for stronger external
fields is not known but one can
expect that for strong enough fields the situation is similar to the
vacuum case.

Though the propagating mode is screened we find that in the static
limit $k_0=0$ the expectation value of the polarization tensor vanishes
(see \eq{wPiw})
which would at first indicate that the instability persists even at finite
temperature. It should however be noticed that an unstable mode is not
static, and as soon as $k_0\neq 0$ the polarization tensor has both a real
part of order $(gT)^2 k_0^2/k^2$, and for $k_0<k$ and imaginary part or order
$(gT)^2 k_0/k$ which are typically much larger than $gB$.
Also, the spectral weight for the lowest Landau level around $k_0=0$ is very
small. We can therefore
conclude that any potentially increasing uniform mode would rapidly be
Landau damped and disappear.
In more realistic models with Higgs field and
fermions the situation is more complicated. In the low temperature phase of
the electroweak theory field strength larger than $m_W^2/e$ is needed to
have an instability \cite{AmbjornO90} but the situation has not been studied
carefully including thermal damping effects. We expect the results in this
paper to give a good picture of what happens in the high temperature phase.
\vskip 5mm
%
%
\acknowledgments
The authors thank NorFA for providing partial financial support under
from the NorFA grant 96.15.053-O. P.~E. was also financially supported by
the Swedish Natural Science Research Council under contract 10542-303.
D.~P.'s research was funded by STINT under contract 97/121.
%
%

%
\end{document}